\documentclass[fleqn,twoside]{article}
\usepackage{espcrc2}
\usepackage[figuresright]{rotating}
\usepackage{graphicx}

\newcommand{\AmS}{{\protect\the\textfont2
  A\kern-.1667em\lower.5ex\hbox{M}\kern-.125emS}}

\newcommand{\intvecx}{\int d^3 x\,}

\newcommand{\vac}{|0\rangle}

\newcommand{\vecp}{{\bf p}}

\newcommand{\vecx}{{\bf x}}       %(vette x)

\newcommand{\al}{\alpha}
\newcommand{\bt}{\beta}
\newcommand{\gm}{\gamma}

\newcommand{\ep}{\epsilon}

\newcommand{\kp}{\kappa}
\newcommand{\lm}{\lambda}
\newcommand{\rh}{\rho}
\newcommand{\sg}{\sigma}

\newcommand{\ph}{\phi}

\newcommand{\om}{\omega}

\newcommand{\half}{\frac{1}{2}}

\newcommand{\tr}{\mbox{tr}\,}

\newcommand{\dmu}{\partial_{\mu}}

\newcommand{\phd}{\ph^{\dagger}}

\newcommand{\eela}[1]{\label{#1}\end{equation}}
\newcommand{\eeala}[1]{\label{#1}\end{eqnarray}}
\newcommand{\be}{\begin{equation}}
\newcommand{\ee}{\end{equation}}
\newcommand{\bea}{\begin{eqnarray}}
\newcommand{\beaa}{\begin{eqnarray*}}
\newcommand{\eea}{\end{eqnarray}}
\newcommand{\eeaa}{\end{eqnarray*}}

\newcommand{\ah}{\hat{a}}
\newcommand{\ahd}{\hat{a}^{\dagger}}

\newcommand{\ncs}{N_{\rm CS}}
\newcommand{\nw}{N_{\rm w}}
\newcommand{\gmsp}{\gm_{\rm sp}}
\newcommand{\ffd}{\tr F\tilde F}
\newcommand{\meff}{m_{\rm eff}}

\title{Classical issues in electroweak baryogenesis}

\author{Jan Smit\address[ITFA]{Institute for Theoretical Physics,
        University of Amsterdam,\\ Valckenierstraat 65, 1018 XE Amsterdam,
        the Netherlands}%
        \thanks{Supported by FOM/NWO}
        \thanks{Presented by J.\ Smit} and
        Anders Tranberg\addressmark[ITFA]}
\begin{document}

\begin{abstract}
In one scenario of baryogenesis, the matter-antimatter asymmetry
was generated in the early universe during a cold electroweak transition.
We model this transition by changing the sign of the effective
mass-squared parameter of the Higgs field from positive
to negative. The resulting `tachyonic' instability leads to a rapid growth of
occupation numbers, such that a classical approximation can be made in
computing subsequent developments in real time.

We solve the classical equations of motion in the SU(2)-Higgs model
under the influence of effective CP-violation. The resulting baryon
asymmetry follows from the generated Chern-Simons number using the
anomaly equation. 
The `classical' difficulties with lattice implementations of these observables
are avoided here because the fields are smooth on the lattice scale.
\vspace{1pc}
\end{abstract}

\maketitle

\section{Introduction}

A basic ingredient in the theories of baryogenesis
is the anomaly in the baryon-current divergence,
\[
\dmu j^{\mu}_B = 3\, q + \cdots,
\;\;\;\;\;\;\;\;
q=\frac{1}{16\pi^2}\, \tr F_{\mu\nu}\tilde F^{\mu\nu},
\]
where
$q$ is the topological-charge density in the SU(2) gauge fields,
which is the divergence of the Chern--Simons current,
$q=\dmu j^{\mu}_{\rm CS}$
(the $\cdots$ denote the U(1) contribution which
is supposed to be inessential).
In electroweak baryogenesis the baryon number $B$ is
generated during the electroweak transition and given by
%\beaa
%B(t) &=& 3\int_0^t dt'\, \intvecx \langle q(\vecx,t')\rangle \\
%&=& 3\langle\ncs(t) - \ncs(0)\rangle,
%\eeaa
\[
B(t) = 3\int_0^t dt'\, \intvecx \langle q(\vecx,t')\rangle
= 3\langle\ncs(t)\rangle,
\]
where $\ncs = \intvecx j^0_{\rm CS}$ is the Chern-Simons number.
We assumed the transition to start at $t=0$ and set $\ncs(0)=0$.

The computation of  $q$ is a `classical' problem in lattice gauge
theory. The `naive' lattice transcription
\[
\tr F(x)\tilde F(x) \propto \ep^{\kp\lm\mu\nu} \tr U_{\kp\lm}(x)
U_{\mu\nu}(x)_{|\rm symmetrized}
\]
suffers from short-distance fluctuations.
Out of equilibrium the problems are worse in the quantum world.

Fortunately, the classical approximation can be used in case of
large occupation numbers, e.g.\ at sufficiently high temperature.
But even in this case there are short-distance problems because of
the slow decrease of occupation numbers at large momentum, $n_p \sim T/p$. 
However, if we have good reason to start with suppressed
short-distance modes, then the problem may show up only at late
times, because classical equipartition is a slow process. For
times of interest to the problem we can then use a `naive' lattice
transcription of $q$. An ideal case is the scenario of
baryogenesis from a rapid cold electroweak transition.

%\section{Scenario (see \cite{ST} and references therein)}
\section{Scenario \cite{Garcia-Bellido:1999sv,ST}}
Suppose the universe is left cold
at the end of low-scale inflation. Subsequently the effective mass
$\meff^2$ in the Higgs potential
\[
\mu^2_{\rm eff} \phd\ph + \lm(\phd\ph)^2
\]
changes sign, e.g.\ due to a coupling
to the inflaton field.
% $\sg$.
% that is increasing towards its vacuum
%expectation value $v_{\sg}$:
%\beaa
%\meff^2 &=& m_{\ph}^2 - \lm_{\sg\ph} \sg^2 >0 \\
%&\to& -\mu^2 = m_{\ph}^2 - \lm_{\sg\ph} v_{\sg}^2<0.
%\eeaa
%$\meff^2 = m_{\ph}^2 - \lm_{\sg\ph} \sg^2 >0
%\to -\mu^2 = m_{\ph}^2 - \lm_{\sg\ph} v_{\sg}^2<0$.
The negative $\meff^2$ causes a (`tachyonic' or `spinodal')
instability in which the system is far out of equilibrium, and
under influence of a CP bias one expects a net change in the
Chern-Simons number, with a corresponding baryon asymmetry.
Non-linear effects lead to rapid approximate thermalization to an
effective temperature $T$ that is expected (counting d.o.f.) to be
much lower than the standard finite-temperature transition
temperature
$T_c$, so subsequent sphaleron transitions are expected to be
negligible.

The reason for the change of sign of $\meff^2$ is uncertain. Not
committing ourselves to any specific cause, we use a sudden
quench:
\beaa
\mu^2_{\rm eff}&=& +\mu^2,\;\;\;\;\;\; t<0,\\
&=& -\mu^2,\;\;\;\;\;\; t>0.
\eeaa
This maximal non-equilibrium setup presumably also maximizes the
asymmetry for given CP violation.

\section{Classical approximation}
The effect of the quench can be studied by first neglecting all
interactions. Let $\hat\ph$ be a real component of the Higgs field
operator. Its spatial Fourier transform behaves around time zero as
\bea
\hat\phi_{\vecp}&=&\frac{1}{\sqrt{2\omega_{p}^{+}}}
\left(\hat a_{\vecp} e^{-i\omega_{p}^{+}t}
+ \hat a_{-\vecp}^{\dagger}e^{i\omega_{p}^{+}t}\right)
,\;\;\; t<0,
\nonumber\\
&=& \hat\al_{\vecp} e^{-i\omega_{p}^{-}t}
+ \hat\bt_\vecp e^{i\omega_{p}^{-}t}
, \;\;\;\;\;\;\;\;\;\;\;\;\;\;\;\;\;\;\;\;\; t >0,
\nonumber\\
%&&\nonumber\\
\omega_{p}^{\pm}&=&\sqrt{\pm\mu^2 +p^{2}}.
\nonumber
\eea
Matching  at $t=0$ determines $\hat\al$ and $\hat\bt$ in terms of
$\ah$ and $\ahd$. The modes $p < \mu$ are unstable and grow exponentially:
%\[\om^-_p = i |\om^-_p|, \;\;\;\;\;
%|\om^-_p| = \sqrt{\mu^2 - p^2}
%\]
$\om^-_p = i |\om^-_p| =i \sqrt{\mu^2 - p^2}$,
\[
\ph_\vecp \to \al_\vecp\, e^{|\om^-_p|t}, \;\;\;\;\;
\pi_\vecp \to |\om^-_p|\, \al_\vecp\, e^{|\om^-_p|t}\approx |\om^-_p| \ph_\vecp
\]
So there is classical behavior for $|\om^-_p| t \gg 1$. Indeed,
generic quantum correlation functions in the initial state $\vac$
(annihilated by the $\ah$) can be well approximated by a classical
distribution \cite{ST},
\[
\exp\left[-\half\sum_{|\vecp|<\mu }
\left(\frac{|\xi^+_\vecp|^2}{n_p + 1/2 + \tilde n_p}
+ \frac{|\xi^-_\vecp|^2}{n_p + 1/2 - \tilde n_p}\right)\right],
\]
%\[
%\xi^{\pm}_{\vecp} =
%\frac{1}{\sqrt{2\om_p}}\left(\om_p\,\ph_{\vecp} \pm
%\pi_\vecp\right),
%\xi^{\pm}_{\vecp} = \left(\om_p\,\ph_{\vecp} \pm
%\pi_\vecp\right)/\sqrt{2\om_p},
%\]
where 
$\xi^{\pm}_{\vecp} = \left(\om_p\,\ph_{\vecp} \pm
\pi_\vecp\right)/\sqrt{2\om_p}$, and
$n$ and $\tilde n$ are instantaneous particle numbers. In
the unstable region, as $\mu t \gg 1$ \cite{ST}:
\[
n_p + 1/2 + \tilde n_p \gg  1,
\;\;\;\;\;
n_p + 1/2 - \tilde n_p \to 0,
\]
so $\xi_\vecp^+$ grows large whereas $\xi_\vecp^- \to 0$. The
canonical coordinates and momenta grow large and become
correlated, an ideal case for making a classical approximation.

As soon as the unstable modes have grown large enough, but before
non-linear interaction effects are expected to take over, we could
sample the above distribution to provide initial conditions for
subsequent classical evolution. However, we may as well sample the
distribution at time zero, since this leads to exactly the same
distribution later on in the free-field approximation. As $n
= \tilde n = 0$ at $t=0$, only the 1/2 remain in the denominators,
which is why we call these `just a half' initial conditions
\cite{ST}. Note that only momenta with $p < \mu$ are initialized.

%As non-linear effects take over the occupation numbers will saturate and
%be redistributed towards effective thermalization.

\section{Simulation}
To study the emerging asymmetry we used the SU(2) Higgs model
with effective CP violation, given by
\bea
-{\cal L} &=& \frac{1}{2g^2}\, \tr F_{\mu\nu} F^{\mu\nu}
+ D_{\mu}\ph^{\dagger} D^{\mu}\ph
\nonumber\\&&\mbox{}
+\frac{\mu^4}{4\lm} -
\mu^2\phd\ph + \lm(\phd\ph)^2
\nonumber\\ && \mbox{}
%+ \kp\, \phd\ph\, \tr F_{\mu\nu}\tilde F^{\mu\nu}
+ \kp\, \phd\ph\, \frac{1}{2} \, \ep^{\mu\nu\rh\sg}
\tr F_{\mu\nu} F_{\rh\sg}.
\nonumber
\eea
Here $\kp$ parametrizes the strength of an effective
CP-violating term of dimension six. It is supposed to represent a leading
effect induced by physics beyond the standard model, or even mimic
CP violation in the standard model itself.
%We also studied an analog model in 1+1 dimensions \cite{ST}.

We discretized the action on a (lorentzian) space-time lattice,
from which the field equations follow in the usual way. Initial
conditions were chosen according to the `just a half' scheme, and
also using a thermal Bose-Einstein distribution with small
$T=0.1\, m_H$ for comparison. Since the initial field
configurations are smooth on the lattice scale, we used `naive'
lattice transcriptions, even for $\ffd$.

The results to follow where obtained with a lattice spacing $am_H=
0.35$, volume $(L m_H)^3 = (60 \times 0.35)^3 = 21^3$, couplings
$g=2/3$ and $\lm$ such that $m_H/m_W = 1$, $\sqrt{2}$ and 2
%We recall the usual expressions for the Higgs and W particle
%masses,
(recall $m_H = \sqrt{2\lm}\, v = \sqrt{2}\, \mu$, $m_W = g v/2$,
with
$v=\mu/\sqrt{\lm}$ the vacuum expectation value of the Higgs field).

\begin{figure}[b]
\vspace{-0.5cm}
%\framebox[55mm]{\rule[-21mm]{0mm}{43mm}}
\includegraphics[width=60mm,clip]{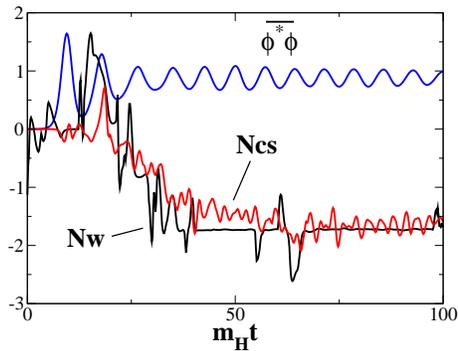}
\vspace{-0.7cm}
\caption{Example of $\phd\ph/v^2$, $\nw$ and $\ncs$ versus
time; $m_H/m_W=1$, thermal initial condition.}
\label{f1}
\end{figure}

Figure \ref{f1} shows the behavior in time of the volume-averaged
$\phd\ph$, the Chern-Simons number and the winding number in the
Higgs field, $\nw$, for one initial condition, for $\kp=0$.
%\vspace{-0.5cm}
%
At first $\phd\ph$ increases exponentially, then peaks at $m_H t =
7.5$ and subsequently executes a damped oscillation towards a
value somewhat smaller than $v^2$. For small energies, in
equilibrium, the winding number is expected to be near $\ncs$, as
this diminishes the covariant derivative $D_{\mu}\ph$. Fig.\
\ref{f1} shows that
$\nw$ behaves at first erratically and then settles near an
integer (the smaller the lattice spacing, the closer to an
integer). The early $\ncs$ is very small and subsequently it
appears to approach
$\nw$. We see no sphaleron transitions (jumps in
$\ncs$ by approximately an integer) in $50 < m_H t < 500$.
The effect of non-zero $\kp$ on single trajectories is
unpredictable, due to the chaotic nature of the dynamics, but an
effect {\em is} present in the average over initial conditions (as
needed for the quantum expectation value).

\begin{figure}[t]
%\vspace{-0.5cm}
%\framebox[55mm]{\rule[-21mm]{0mm}{43mm}}
\includegraphics[width=60mm,clip]{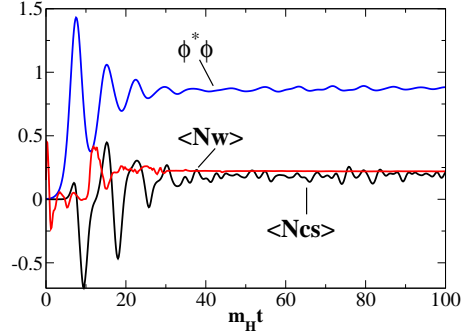}
\vspace{-0.7cm}
\caption{$\langle \phd\ph/v^2\rangle$, $\langle\ncs\rangle$ and $\langle\nw\rangle$
for $\kp = (1/2\pi^{2}) m_W^{-2}$, $m_H/m_W=\sqrt{2}$, and `just a
half' initial conditions.}
\label{f2}
\end{figure}
%\vspace{-0.5cm}
%
Figure \ref{f2} shows an example of averages for non-zero 
$\kp$. After an initial
rise of $\langle\ncs\rangle$ (which we understand semi
analytically), there appears to be a sort of resonance with
$\langle \phd\ph\rangle$ that may even lead (for larger $\kp$)
to a
$\langle\ncs\rangle$ of opposite sign.
% as expected from the CP bias.
We found a similar resonant effect in 1+1 dimensions, where it
leads to a strong dependence on $m_H/m_W$ \cite{ST}.
Note that after $t m_H = 40$ there is no visible drift of 
$\langle\ncs\rangle$ towards zero due to sphaleron transitions.

\section{Conclusion}
Baryogenesis through tachyonic electroweak transition offers ideal
case for lattice simulations using the classical approximation.
Together with Jon-Ivar Skullerud we have also computed Higgs and W
particle numbers \cite{SST}. As expected, they grow large
at low momenta, while staying negligible for momenta $p > 1.5\,
m_H$. This supports our experience that the
effective CP-violating interaction
$-{\cal L}_{\rm CP} =  \kp \int \phd\ph\, \ffd$
can be meaningfully implemented using a `naive' lattice
transcription. 
%See \cite{ST2} for more details.
More details are in \cite{ST2} (and to be published).

The resulting asymmetry appears to be substantial: using simple
estimates the observed baryon to photon ratio can be reproduced
with
$\kp \approx 1\times 10^{-5}\,\mbox{TeV}^{-2}$, assuming $m_H =
\sqrt{2}\, m_W$.
%The temperature after the transition should be low enough for $B$
%not to be washed out by sphaleron processes.
% ($\dot B \propto -\gmsp\, B/T^3$, with $\gmsp$ the sphaleron rate).


\begin{thebibliography}{9}
%\cite{Garcia-Bellido:1999sv}
\bibitem{Garcia-Bellido:1999sv}
J.\ Garc\'{\i}a-Bellido, D.Y.\ Grigoriev, A.\ Kusenko and M.E.\
Shaposhnikov,
%``Non-equilibrium electroweak baryogenesis from preheating after  inflation,''
Phys.\ Rev.\ D {\bf 60} (1999) 123504. %[arXiv:hep-ph/9902449];
%\cite{Krauss:1999ng}
%\bibitem{Krauss:1999ng}
L.~M.~Krauss and M.~Trodden,
%%``Baryogenesis below the electroweak scale,''
Phys.\ Rev.\ Lett.\  {\bf 83} (1999) 1502;
% [arXiv:hep-ph/9902420].
%%CITATION = HEP-PH 9902420;%%
%\cite{Copeland:2001qw}
%\bibitem{Copeland:2001qw}
E.J.\ Copeland, D.\ Lyth, A.\ Rajantie and M.\ Trodden,
%``Hybrid inflation and baryogenesis at the TeV scale,''
Phys.\ Rev.\ D {\bf 64} (2001) 043506.
%[arXiv:hep-ph/0103231].
%%CITATION = HEP-PH 0103231;%%
\bibitem{ST} J.\ Smit, A.\ Tranberg, JHEP 0212 (2002) 020.
%[hep-ph/0211243].
\bibitem{SST} J.-I. Skullerud, J.\ Smit and A.\ Tranberg,
JHEP 0212 (2002) 020;
% [hep-ph/0211243];
these proceedings [hep-lat/0309046].
\bibitem{ST2} J.\ Smit, A.\ Tranberg, hep-ph/0210348.
%and to be published.

\end{thebibliography}
\end{document}